\documentclass[prb,twocolumn,showpacs,floatfix,amsmath,amssymb,superscriptaddress]{revtex4-1}

\usepackage{amsfonts}
\usepackage{stmaryrd}
\usepackage{bbm}
\usepackage{mathrsfs}
\usepackage{tipa}
\usepackage{amssymb}
\usepackage{txfonts}
\usepackage{graphicx}
\usepackage{dcolumn}
\usepackage{epstopdf}
\usepackage[colorlinks,linkcolor=blue,urlcolor=blue,citecolor=blue]{hyperref}
\usepackage{multirow}
\usepackage{subfigure}
\usepackage{url}
\usepackage{setspace}

\begin{document}

\title{Optical spectroscopy and band structure calculations of structural phase transition in the Vanadium-based kagome metal ScV$_6$Sn$_6$}%force line break\\

\author{Tianchen Hu}
\email{These authors contributed equally to this work.}
\affiliation{International Center for Quantum Materials, School of Physics, Peking University, Beijing 100871, China}

\author{Hanqi Pi}
\email{These authors contributed equally to this work.}
\affiliation{Beijing National Laboratory for Condensed Matter Physics and Institute of Physics, Chinese Academy of Sciences, Beijing 100190, China}
\affiliation{School of Physical Sciences, University of Chinese Academy of Sciences, Beijing 100049, China}

\author{Shuxiang Xu}
\affiliation{International Center for Quantum Materials, School of Physics, Peking University, Beijing 100871, China}

\author{Li Yue}
\affiliation{International Center for Quantum Materials, School of Physics, Peking University, Beijing 100871, China}

\author{Qiong Wu}
\affiliation{International Center for Quantum Materials, School of Physics, Peking University, Beijing 100871, China}

\author{Qiaomei Liu}
\affiliation{International Center for Quantum Materials, School of Physics, Peking University, Beijing 100871, China}

\author{Sijie Zhang}
\affiliation{International Center for Quantum Materials, School of Physics, Peking University, Beijing 100871, China}

\author{Rongsheng Li}
\affiliation{International Center for Quantum Materials, School of Physics, Peking University, Beijing 100871, China}

\author{Xinyu Zhou}
\affiliation{International Center for Quantum Materials, School of Physics, Peking University, Beijing 100871, China}

\author{Jiayu Yuan}
\affiliation{International Center for Quantum Materials, School of Physics, Peking University, Beijing 100871, China}

\author{Dong Wu}
\affiliation{Beijing Academy of Quantum Information Sciences, Beijing 100913, China}

\author{Tao Dong}
\affiliation{International Center for Quantum Materials, School of Physics, Peking University, Beijing 100871, China}

\author{Hongming Weng}
\email{hmweng@iphy.ac.cn}
\affiliation{Beijing National Laboratory for Condensed Matter Physics and Institute of Physics, Chinese Academy of Sciences, Beijing 100190, China}
\affiliation{School of Physical Sciences, University of Chinese Academy of Sciences, Beijing 100049, China}

\author{Nanlin Wang}
\email{nlwang@pku.edu.cn}
\affiliation{International Center for Quantum Materials, School of Physics, Peking University, Beijing 100871, China}
\affiliation{Beijing Academy of Quantum Information Sciences, Beijing 100913, China}
\affiliation{Collaborative Innovation Center of Quantum Matter, Beijing 100871, China}

\begin{abstract}

 In condensed matter physics, materials with kagome lattice display a range of exotic quantum states, including charge density wave (CDW), superconductivity and magnetism. Recently, the intermetallic kagome metal ScV$_6$Sn$_6$ was discovered to undergo a first-order structural phase transition with the formation of a $\sqrt{3}$$\times$$\sqrt{3}$$\times$ 3 CDW at around 92 K. The bulk electronic band properties are crucial to understanding the origin of the structural phase transition. Here, we conducted an optical spectroscopy study in combination with band structure calculations across the structural transition. Our findings showed abrupt changes in the optical reflectivity/conductivity spectra as a result of the structural transition, without any observable gap formation behavior. The optical measurements and band calculations actually reveal a sudden change of the band structure after transition. It is important to note that this phase transition is of the first-order type, which distinguishes it from conventional density-wave type condensations. Our results provide an insight into the origin of the structural phase transition in this new and unique kagome lattice intermetallic.
\end{abstract}

\pacs{Valid PACS appear here}

\maketitle

\section{INTRODUCTION}

The unique kagome lattice is a two-dimensional network of corner-sharing triangles which have gained tremendous interests for studying the latent interplay of frustrated, correlated and topological nontrivial electronic states\cite{Ye2018,Yin2019,Liu2020,Yin2020,Kang2020,Yin2022}. Tight-binding models suggest that the electronic structure could host Dirac nodes, van Hove singularities and geometrically driven flat bands in kagome lattice\cite{Guo2009,Beugeling2012}. And the fertile electronic ground states of kagome lattice systems could be superconductivity (SC)\cite{Ortiz2020}, charge density waves (CDW)\cite{Ortiz2020,Ortiz2019,Teng2022}, spin density waves\cite{Yu2012} or a quantum spin liquid\cite{Balents2010,Yan2011}, etc.

The intriguing characteristics of the CDW and its delicate interactions with superconductivity have been extensively studied in various condensed matter systems. The conventional CDW is often attributed to Fermi surface nesting (FSN) and many of these CDW-bearing materials are also superconducting\cite{Morosan2006,Sakamoto2007,Ortiz2020} . The unconventional SC and CDW were discovered to coexist in the new correlated kagome metal AV$_3$Sb$_5$ (A= K, Rb, Cs) systems before\cite{Ortiz2020}, immediately sparking much interest in the novel physics involved in these materials\cite{Chen2021a,Chen2021,Zhao2021,Liang2021,Yu2021}. Currently, the origin of CDW and SC is a topic of ongoing debates among AV$_3$Sb$_5$. For instance, stacked Vanadium kagome layers and antimony-bands have been proposed to play an important role \cite{Oey2022,Nakayama2022,Tsirlin2022}.The CDW formation is also closely tied to the electron-phonon coupling and the electronic band saddle point nesting\cite{Hu2022a,Zhou2021,PhysRevB.105.L140501}. Recently, new Vanadium-based kagome metal ScV$_6$Sn$_6$ of a large HfFe$_6$Ge$_6$-type (space group: No.191, {\itshape{P}}6/mmm at 300 K) family was discovered to undergo a first-order phase transition at around T$_s$$\approx$92 K, reminiscent of the popular CsV$_3$Sb$_5$ system\cite{Arachchige2022}, but no SC has been observed at low temperatures. X-ray diffraction has verified that the formation of a three-dimensional $\sqrt{3}\times\sqrt{3}\times$ 3 periodic CDW modulation below T$_s$. And high-pressure transport study have shown that the CDW order can be completely suppressed but without SC emerging, even under pressures up to 11 GPa\cite{Zhang2022}. Unlike AV$_3$Sb$_5$ systems, ScV$_6$Sn$_6$ has two kagome sheets per unit cell separated by alternating ScSn$_2$ and Sn$_2$ layers in ScV$_6$Sn$_6$. Among RV$_6$Sn$_6$ (R =Y, Gd-Tm, and Lu) with their intriguing f-orbital magnetism, no Vanadium-driven order has been observed to date\cite{Pokharel2021,Peng2021,Hu2022b,Zhang2022b}. The transport behaviors and the filling of the Vanadium d-orbital bands are similar to AV$_3$Sb$_5$, making ScV$_6$Sn$_6$  an ideal system for further understanding the origin of CDW in kagome lattices. 

\begin{figure*}[htbp]
	\centering
	\includegraphics[width=18cm]{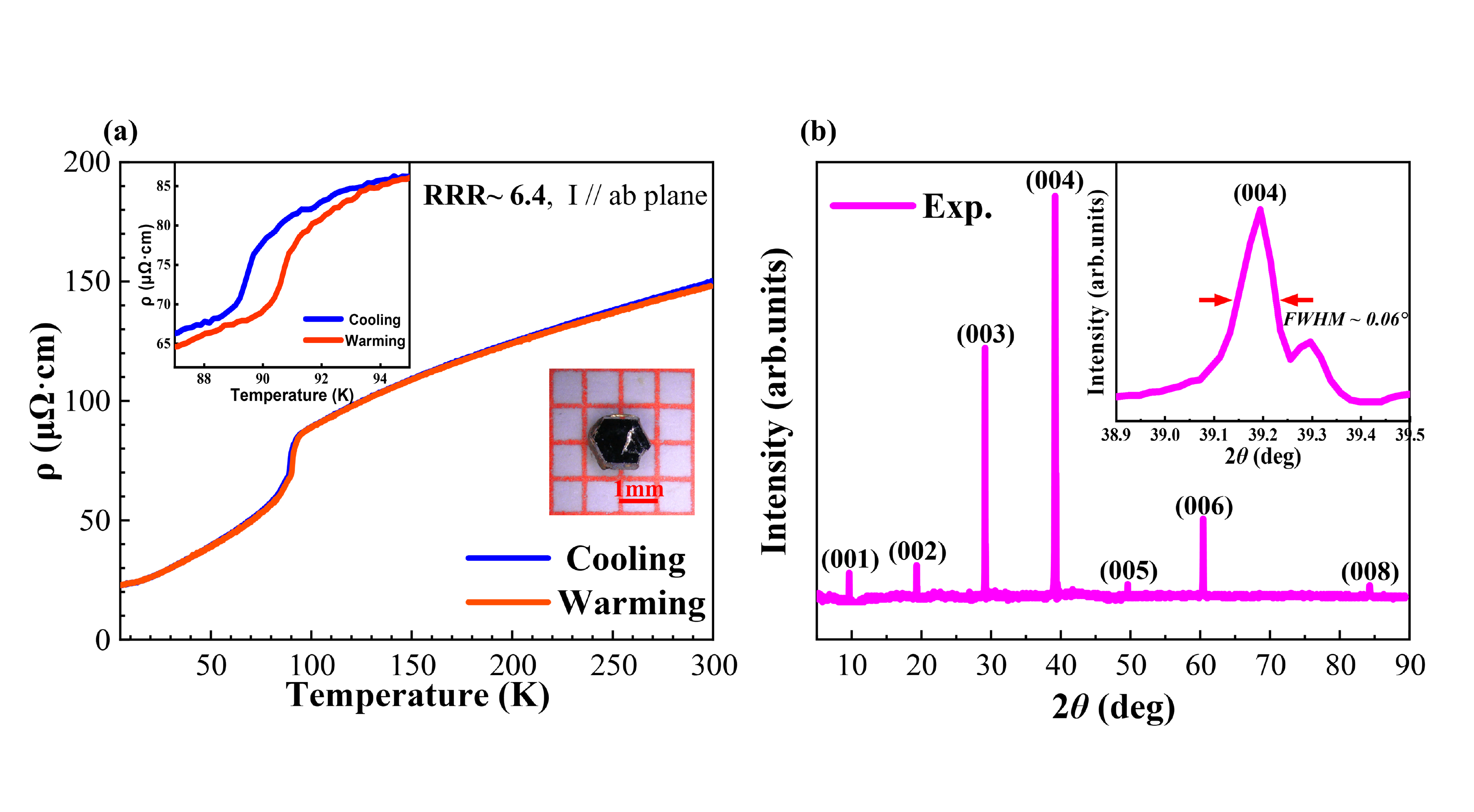}\\
	\caption{\textbf{Sample physical characterizations for ScV$_6$Sn$_6$.} (a) Temperature-dependent ab-plane resistivity measurement, a hysterisis is evident near 92 K. Inset: an optical micrograph of ab plane of ScV$_6$Sn$_6$ single crystal on a 1mm grid (b) Room temperature powder XRD patterns of single-crystal and indexing. Inset: The full width at half maximum of (004) bragg peak is about 0.06$^{\circ}$, indicating good quality of single crystal.
}\label{Fig:1}
\end{figure*}

It is crucial to get insights into the electronic band structure across the phase transition through various spectroscopy techniques. To date, no spectroscopy results have been reported using angle resolved photoemission spectroscopy or scanning tunneling microscope. Optical reflectivity/conductivity spectroscopy is powerful and sensitive for detecting the bulk electronic states of solids in many cases. So it is significant to uncover the underlying physics of the phase transition by optical reflectivity measurements.

In this study, we have successfully grown   and characterized hexagonal single-crystalline ScV$_6$Sn$_6$. 
Our resistivity measurements revealed a pronounced first-order phase transition around 92 K. Similar to some CDW systems\cite{Soergel2006,Ortiz2019}, the compound's dc conductivity is enhanced below T$_s$\cite{Arachchige2022}. To further understand this issue, we performed temperature-dependent optical reflectivity spectroscopy in combination with band structure calculations on ScV$_6$Sn$_6$. Both measurements and band structure calculations reveal a sudden change of band structure, with no evidence of gap development. The sudden changes in the electronic structure are also observed in some first-order structural phase transitions\cite{Fang2013,Chen2015,Hu2022}. Therefore, we could elaborate that the driving mechanism of CDW formation is not the conventional FSN in the ScV$_6$Sn$_6$. Comparing with the well-studied AV$_3$Sb$_5$, the gap-opening behaviors and some first-order phase transition features are both observed. The case seems to be different in ScV$_6$Sn$_6$. Our work provides insight into CDW formation in the new Vanadium-based kagome metal intermetallic compound.

\section{METHODS}
The single crystals of ScV$_6$Sn$_6$ were synthesized using the Sn self-flux method with a 1:6:40 atomic ratio of Sc:V:Sn. High-purity scandium grains (99.99\%, 1-5 mm), vanadium powders (99.9\%, 200 mesh), and stannum shots (99.999\%, 1-3 mm) were placed in an aluminium crucible and sealed in a fused silica tube under high vacuum. The quartz tube was slowly heated to 1150 $^{\circ}$C in a furnace followed by a 20h dwell and then cooled to 750 $^{\circ}$C at a rate of 1 $^{\circ}$C/h. After removing the Sn flux through high-speed centrifuging, single crystals of ScV$_6$Sn$_6$ with hexagonal shiny facet (typical size of 2$\times$2$\times$0.5 mm$^3$, as shown in the inset of Fig.\ref{Fig:1} (a)) were obtained. 

The temperature-dependent resistivity was measured using a four-probe method with the current direction along the ab plane in a Quantum Design physical property measurement system. The resistivity showed good metallic behavior with a value of about 150 $\mu\Omega\cdot$cm at 300 K and a residual resistance ratio of 6.4 (shown in Figure \ref{Fig:1}(a)). Similar to the AV$_3$Sb$_5$ (A= K,Rb,Cs) CDW systems\cite{Ortiz2019}, ScV$_6$Sn$_6$ remains metallic state at low temperature (LT) phase. The resistivity curve also shows a clear hysteresis at around 92 K (as shown in  \ref{Fig:1}(a)), indicating a first-order type phase transition. X-ray diffraction results at ambient conditions are shown in Fig.\ref{Fig:1} (b). The seven distinct diffraction peaks of  Fig.\ref{Fig:1} (b) can be indexed as (00L) of ScV$_6$Sn$_6$ (deriving from CIF files in literature\cite{Arachchige2022}). The high quality of the ScV$_6$Sn$_6$ single crystals was confirmed by the small full width about 0.06$^{\circ}$ at half maximum of the strongest (004) Bragg peak in the inset of Fig.\ref{Fig:1} (b). Chemical compositions measurements by energy dispersive spectroscopy (EDS) showed that the atomic ratios were close to the standard 1:6:6 stoichiometric ratio (Sc:V:Sn $\sim$ 1: 5.88 : 6.27). All the sample fundamental characterizations are identical with the literature.
\cite{Arachchige2022,Zhang2022}.
\begin{figure*}[htbp]
	\centering
	\includegraphics[width=18cm]{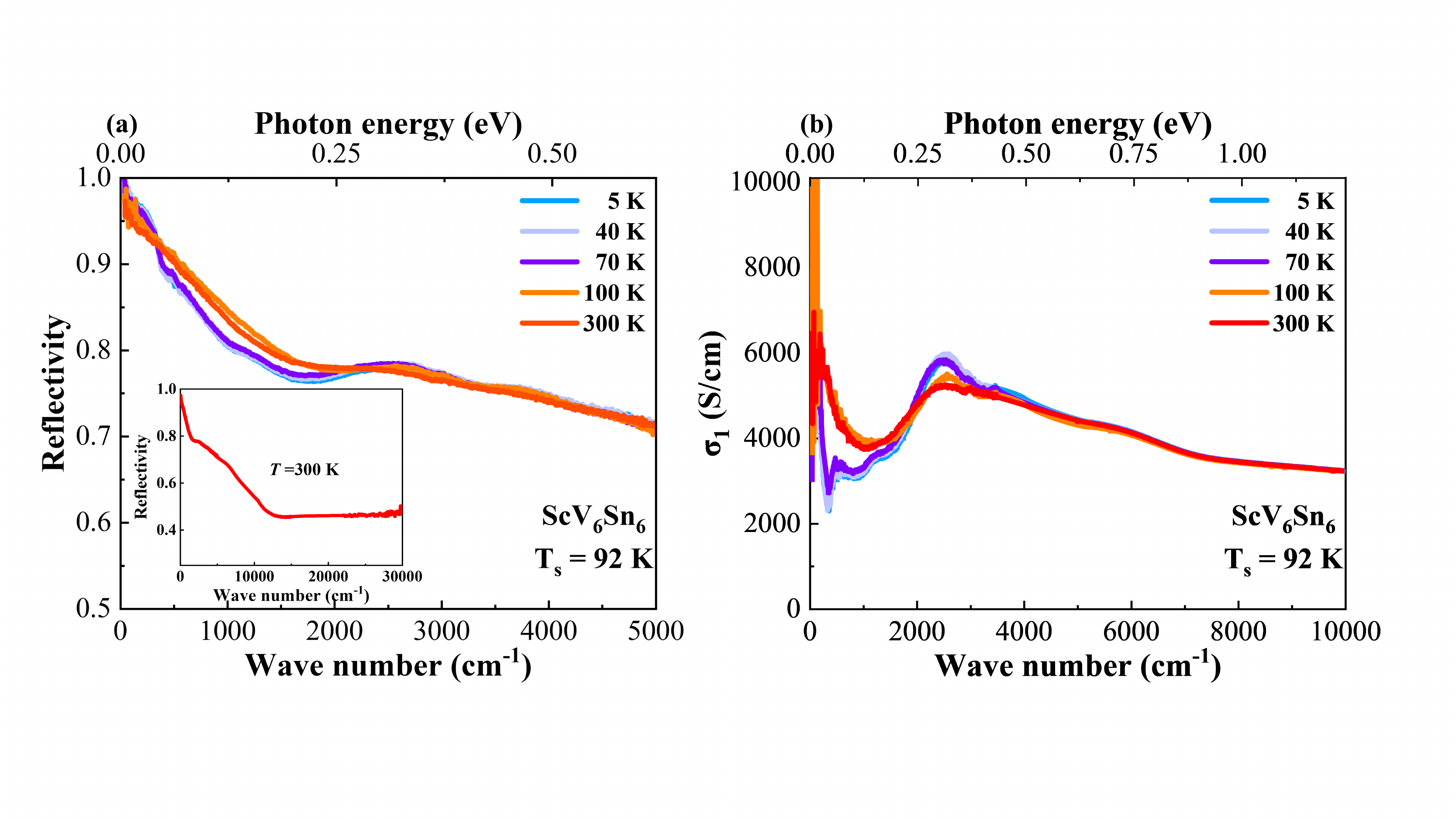}\\
	\caption{\textbf{Temperature-dependent optical reflectivity/conductivity of ScV$_6$Sn$_6$ } (a) temperature-dependent optical reflectivity measurements below 5000 cm$^{-1}$, Inset: Large energy scale range of 50-30000 cm$^{-1}$ at 300 K. (b) temperature-dependent optical conductivity below 10000 cm$^{-1}$.
}\label{Fig:2}
\end{figure*}

The hexagonal ab-plane optical reflectance measurements were performed on a Fourier transform infrared spectrometer (Bruker Vertex 80V) at near-normal incidence with a frequency range from 40 to 30000 cm$^{-1}$. The sample is attached to a copper cone in order to reduce stray light influence. And an in-situ gold and aluminum evaporation coating technique is performed at 300 K with purpose of getting the absolute reflectance $R(\omega)$. The temperature-dependent optical reflectivity data were collected as the sample was warmed up to the target temperature. 

The optical conductivity spectra were derived from the Kramers-Kronig transformation of $R(\omega)$. We use Hagen-Rubens' relation for the low-frequency extrapolation. For the high-frequency side, we extrapolate the measured reflectance constantly up to 100000 cm$^{-1}$, then connected to the X-ray atomic scattering functions\cite{Tanner2015}. 

The electronic structures of ScV$_6$Sn$_6$ were calculated using the Vienna ab initio simulation package (VASP) \cite{Kresse1996} with the generalized gradient approximation of Perdew-Burke-Ernzerhof exchange-correlation potential \cite{PhysRevLett.78.1396}. The self-consistent calculation of the HT phase and the LT phase was carried out on an 11×11×6 k-mesh and a 7×7×7 k-mesh, respectively, with the energy cutoff of 500 eV. The intraband part is simulated with the Drude model, 
\begin{equation}
\sigma_{\text {Drude }}(\omega)=\frac{\omega}{4 \pi} \operatorname{Im} \epsilon_{\text {Drude }}(\omega)=\frac{1}{4 \pi} \frac{\omega_p^2 \gamma}{\omega^2+\gamma^2}
\end{equation}
where $\omega_p$ is the plasma frequency, and $\gamma$ is the carrier scattering rate.
The interband optical conductivity $\sigma_1$($\omega$) is computed with the Kubo-Greenwood formula,
\begin{equation}
\sigma_{\text {inter } ; \alpha \beta}=\frac{i e^2 \hbar}{N_k \Omega_c} \sum_k \sigma_{\alpha \beta, k}(\omega)
\end{equation}
\begin{equation}
=\frac{i e^2 \hbar}{N_k \Omega_c} \sum_k \sum_{n, m} \frac{f_{m k}-f_{n k}}{\varepsilon_{m k}-\varepsilon_{n k}} \frac{\left\langle\psi_{n k}\left|v_\alpha\right| \psi_{m k}\right\rangle\left\langle\psi_{m k}\left|v_\beta\right| \psi_{n k}\right\rangle}{\varepsilon_{m k}-\varepsilon_{n k}-(\hbar \omega+i \eta)}
\end{equation}
where $\alpha$, $\beta$ denote Cartesian directions, $\Omega_c$ is the cell volume, $\textit N_k$ denotes the number of k-points sampling the BZ, $\varepsilon_{mk}$ is the band energy and $\textit f_{mk}$ is the Fermi-Dirac distribution function. The k-meshes in the optical calculation were 30×30×20 and 20×20×20 in the HT phase and the LT phase, respectively. Besides, the gaussian smearing factors are larger than the scattering rate obtained from the Drude model. The Fermi surface was calculated using the WANNIERTOOLS package \cite{Wu2018} on the 61×61×61 k-mesh. A 100×100×100 k-mesh and 50×50×1 q-mesh were used to obtain the electron susceptibility.

\section{RESULTS AND DISCUSSION}
\begin{figure*}[htbp]
	\centering
	\includegraphics[width=18cm]{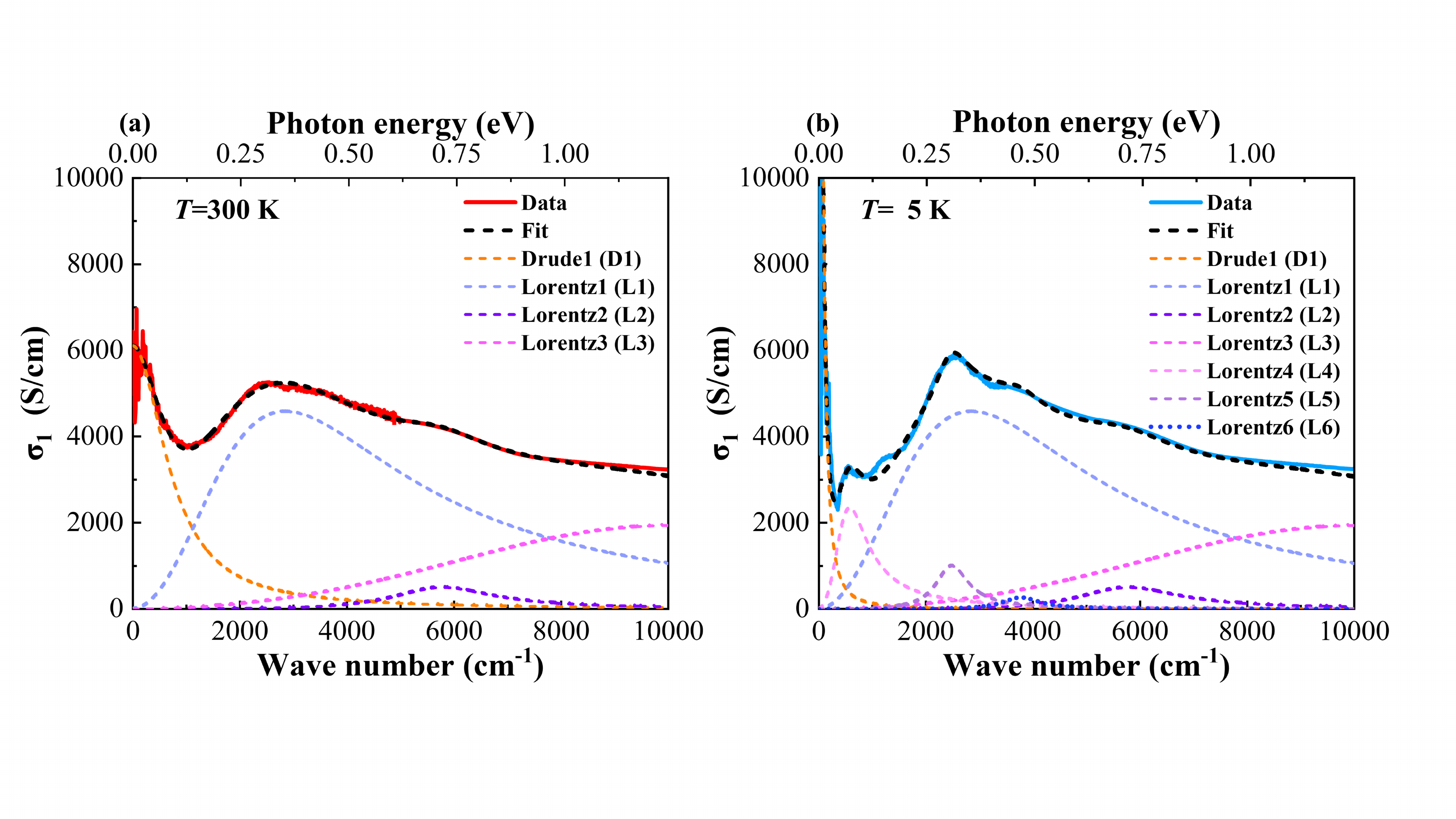}\\
	\caption{\textbf{The Drude-Lorentz model fitting results of the optical conductivity for ScV$_6$Sn$_6$.} (a) $T$=300 K. (b) $T$=5 K.
}\label{Fig:3}
\end{figure*}
 The main panel of Fig.\ref{Fig:2} (a) shows the reflectivity up to 5000 cm$^{-1}$ at five different temperatures. The inset displays the experimental reflectance spectrum up to 30000 cm$^{-1}$ at 300 K. In the low frequency region, $R(\omega)$ has high values and approaches unity at zero-frequency limit, reflecting the metallic nature of the compound in the spectra.  In the high temperature phase, there is only a minor change in $R(\omega)$ between 100 K and 300 K. A roughly linear frequency dependence is observed below 2000 cm$^{-1}$ in $R(\omega)$. The spectral shape reveals the strong damping behavior of charge carriers, suggesting that the charge carriers experience strong scattering. However, when the temperature drops just below T$_s$, the optical reflectivity $R(\omega)$ shows distinct differences up to 5000 cm$^{-1}$. Further decreases in temperature result in minimal changes to the spectral features. Below T$_s$, in the far-infrared region, $R(\omega)$ increases slightly and shows an edge-like shape near 500 cm$^{-1}$, indicating a sudden reduction of plasma frequency with much reduced carrier scattering rate. Above the edge, the reflectance is connected to a linear frequency dependent behavior. Additionally, Several weak peak-like features appear in $R(\omega)$ near 2460 cm$^{-1}$, 3780 cm$^{-1}$.

Figure 2(b) presents the real part of optical conductivity $\sigma_1(\omega)$  below 10000 cm$^{-1}$ calculated from the $R(\omega)$ data using the Kramers-Kronig transformation. The Hagen-Rubens relation and  x-ray atomic scattering functions were utilized for the low-energy and high-energy extrapolation of $R(\omega)$\cite{Tanner2015}, respectively. The main panel of Fig.\ref{Fig:2} (b) displays $\sigma_1(\omega)$ below 10000 cm$^{-1}$ at five selected temperatures. All spectra exhibit a pronounced Drude response whose peak is centered at zero frequency, which is a typical characteristic of metals. In the HT phase, the broad width of the Drude peak indicates a large scattering rate of the itinerant carriers. Upon entering the LT phase, a portion of Drude-type spectral weight is abruptly transferred to high-energy excitations, leading to the formation of three distinctive peaks at 570 cm$^{-1}$ , 2460 cm$^{-1}$ and 3780 cm$^{-1}$. The overall spectral change just below T$_s$ reflects a significant band structure change behavior associated with the structural phase transition. Unlike in conventional CDW order formation, the overall shape of the conductivity spectra remains almost unchanged at 5 K, 40 K and 70 K.  

As we know, in many conventional CDW systems such as rare-earth tri-telluride (RTe$_3$)\cite{Hu2014}, LaAgSb$_2$\cite{Chen2017}, Bi$_2$Rh$_3$Se$_2$\cite{Lin2020}, it can be clearly observed that the optical spectral continuous change only occurs in the low energy region at low temperatures. The formation of CDW gap is a hallmark of second-order phase transition behavior. The so-called case-\uppercase\expandafter{\romannumeral1} coherence factor of CDW condensate\cite{MD2002Electrodynamics}  causes continuous spectral weight transfer, producing increasing peaks just above the energy gap in the conductivity. As the temperature decreases further from T$_{cdw}$, the suppression features in the reflectivity become more prominent. This is a widely accepted optical characteristic of conventional CDW condensation and possible related structural transition.On the other hand, unlike the continuous second-order  behaviors, there could be unpredictable or even entirely different optical reflectivity spectra below and above the transition for  pure first-order structural phase transition, as previously observed in IrTe$_2$\cite{Fang2013} ,RuP\cite{Chen2015}  and TaTe$_2$\cite{Hu2022} CDW systems before. Additionally, the optical reflectivity line shape remains nearly unchanged with temperature variation below T$_s$. 
In the case of ScV$_6$Sn$_6$, no continuous spectral change was observed upon entering the LT CDW phase. Instead, the spectra below T$_s$ show only minor differences with decreasing temperature. Based on these results and analysis, we can conclude that the transition is of pure first-order nature.

\begin{figure*}[htbp]
	\centering
	\includegraphics[width=16cm]{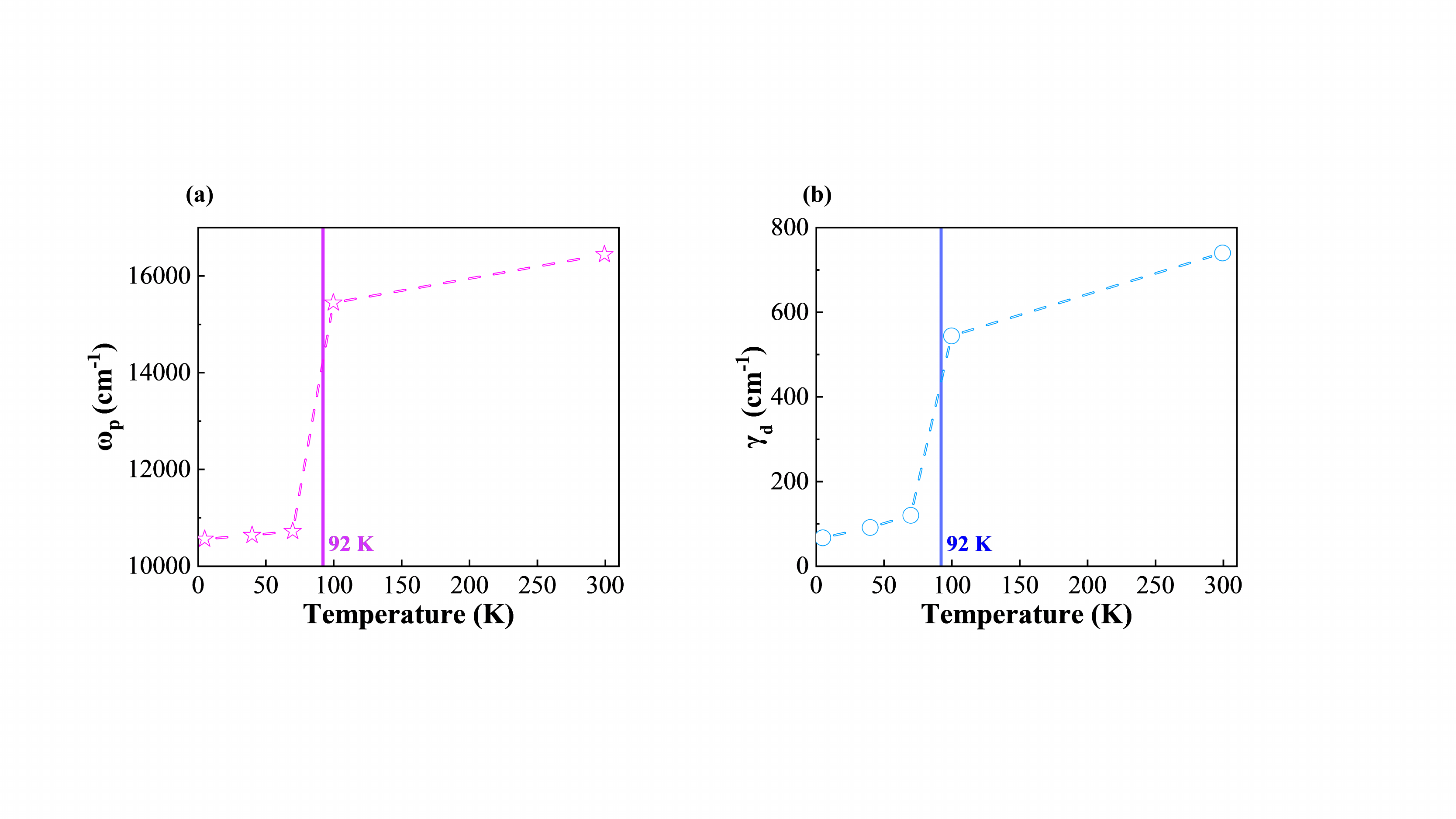}\\
	\caption{\textbf{Values of the temperature-dependent fitting parameters} (a) plasma frequency and (b) scattering rate.
}\label{Fig:4}
\end{figure*}

It should be noted that the measuring reflectivity at 30000 cm$^{-1}$ is a relatively high value of approximately 0.45. This high-energy side extrapolation in Kramers-Kronig transformation could impact the integration results and conductivity values. However, it is worth mentioning that the low-energy conductivity remains nearly unaffected. In this work we are mainly focus on the spectral characteristics and change below and above the transition, so the exact numerical values of the conductivity peaks do not impact our discussions and conclusions. 

To better distinguish and analyze the electronic excitations in optical conductivity data, we employ the classical Drude-Lorentz model to fit the optical conductivity. The model's general formula can be describe as \begin{equation}
\sigma_{1}(\omega)=\sum_i \frac{\omega_{p i}^{2}}{4 \pi} \frac{\Gamma_{D i}}{\omega^{2}+\Gamma_{D i}^{2}}+\sum_j \frac{S_{j}^{2}}{4 \pi} \frac{\Gamma_{j} \omega^{2}}{\left(\omega_{j}^{2}-\omega^{2}\right)^{2}+\omega^{2} \Gamma_{j}^{2}}
\end{equation}
where $\omega_{p i}$ and $\Gamma_{D i}$ are the plasma frequency and the scattering rate of each conduction band while $\omega_{j}$, $\Gamma_{j}$, and $S_{j}$ represents resonance frequency, the damping, and the mode magnitude of each Lorentz oscillator, respectively. The first term is Drude component, representing the contributions from the free conduction carriers. The second Lorentz component terms are used to describe the excitations across energy gaps. 
In our study, we found that only one Drude component was sufficient to approximate the low frequency conductivity in ScV$_6$Sn$_6$, which was unexpected based on previous findings in kagome metals\cite{Zhou2021,Uykur2021,Biswas2020,Wenzel2022,Wenzel2022a,Uykur2022}. The optical conductivity is well represented by a Drude component (D$_1$) and three Lorentz oscillators (L$_1$, L$_2$ and L$_3$) at 300 K. However, below the transition temperature, the Drude component became substantially narrowed and three additional Lorentz oscillators ( L$_4$,L$_5$ and L$_6$) are required to fit the data. Fig.\ref{Fig:3} (a) and (b) present the experimental date of $\sigma_1(\omega)$ at 5 K and 300 K respectively, with the black dashed line representing the sum of the Drude-Lorentz fitting. In addition, the Lorentz 3 stems from the high-energy interband transition, which is stable with all temperature variations.
Figure \ref{Fig:4} displays the two fitting parameters of the Drude component. With decreasing the temperature, both the plasma frequency ($\omega_{p}$) and scattering rate ($\gamma_{d}$) decrease, as shown in Figure \ref{Fig:4}(a) and \ref{Fig:4}(b). Since the square of plasma frequency is related to the carrier densities, we can infer a portion of the free carriers are likely to lost after transition, while the sharp drop in scattering rate helps to explain the improved dc conductivity in the LT phase.

\begin{figure*}[htbp]
	\centering
	\includegraphics[width=18cm]{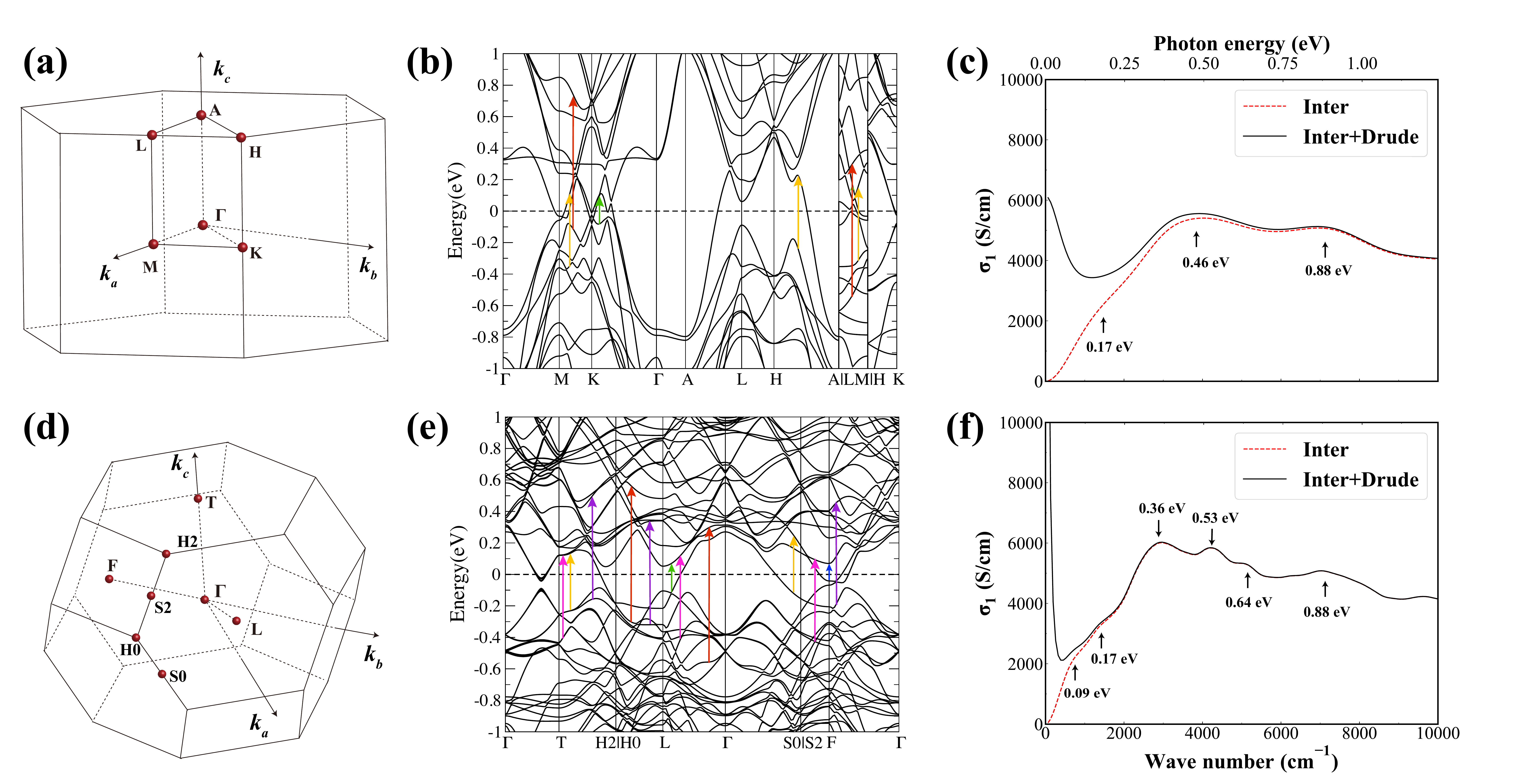}\\
	\caption{\textbf{The band and optical conductivities calculation results for ScV$_6$Sn$_6$.} 
  (a,d) Brillouin zones, (b,e) band structures with spin-orbit coupling (SOC) included and (c,f) calculated $\sigma_1(\omega)$ of ScV$_6$Sn$_6$ in the HT phase (a-c) and LT phase (d-f). The optical conductivities without (with) Drude contribution are presented with red dash lines (black solid lines) in (c,f). The green/yellow/red arrows in (b) denote the transitions at 0.17/0.46/0.88 eV, while the blue/green/yellow/pink/violet/red arrows in (e) denote the transitions at 0.09/0.17/0.36/0.53/0.64/0.88 eV.
}\label{Fig:5}
\end{figure*}

To explore the origin of the components in $\sigma_1 (\omega)$, we also perform the first-principles calculations to obtain the band structures and optical conductivities of ScV$_6$Sn$_6$ as reference. The Brillouin zone (BZ) and band structure with spin-orbit coupling (SOC) included in the HT phase are shown in Fig.\ref{Fig:5}(a) and \ref{Fig:5}(b). The energy bands near the Fermi level, which host the Dirac node, van Hove singularities, and flat bands on the $\Gamma$-M-K-$\Gamma$ path, are mainly contributed by d-orbitals of V atoms occupying the kagome sites. The calculated  $\sigma_1 (\omega)$ with only interband contribution is presented in Fig.\ref{Fig:5}(c) with red dash line and has three absorption peaks at 1370, 3710, and 7100 cm$^{-1}$ ($\approx$0.17, 0.46, and 0.88 eV). We mark the corresponding electron transitions with green, yellow, and red arrows in  Fig.\ref{Fig:5}(b), which connect the nearly parallel bands and thus have a large joint density of states. The calculated plasma frequency is obtained as $\omega_p$=15210 cm$^{-1}$ ($\approx$1.886 eV) and the scattering rate is $\gamma$=630 cm$^{-1}$ ($\approx$0.078 eV) to well fit the experimental spectra with the Drude model. The optical conductivity with Drude contribution included is represented in  Fig.\ref{Fig:5}(c) with black solid lines. The absorption peak at 1370 cm$^{-1}$ is hidden by the response of free charge carriers.  Fig.\ref{Fig:5}(d-e) shows the BZ and band structure with SOC included in the LT phase. Compared to the interband optical response in the HT phase, interband $\sigma_1 (\omega)$ in the LT phase has a new absorption peak at 730 cm$^{-1}$ ($\approx$0.09 eV) as shown in Fig.\ref{Fig:5}(f) with the red dash line. The corresponding electron transition is represented by blue arrows near the van Hove singularities in  Fig.\ref{Fig:5}(e). Moreover, the absorption peak at 3710 cm$^{-1}$ in the HT phase splits into three peaks at 2900, 4270, and 5160 cm$^{-1}$ ($\approx$ 0.36, 0.53, and 0.64 eV) in the LT phase, which are denoted by yellow, pink, and violet arrows in  Fig.\ref{Fig:5}(f), respectively. The optical response with Drude contribution is shown by the black solid line in  Fig.\ref{Fig:5}(f) with the calculated plasma frequency $\omega_p$=11930 cm$^{-1}$ ($\approx$1.48 eV) and the fitting scattering rate $\gamma$=60 cm$^{-1}$ ($\approx$0.008 eV).  
\begin{figure*}[htbp]
	\centering
	\includegraphics[width=18cm]{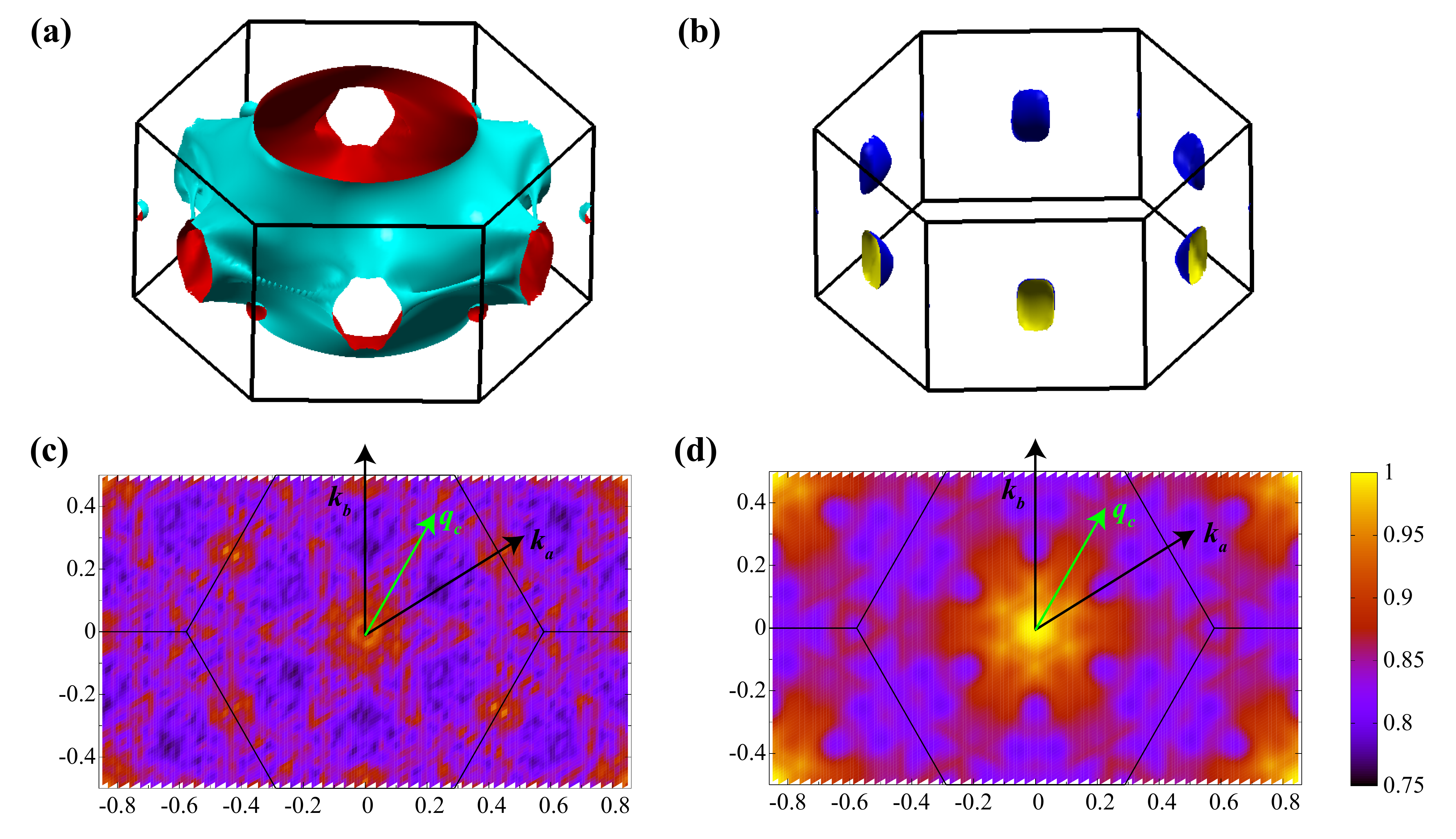}\\
	\caption{\textbf{The calculation Fermi surface and electron susceptibility
of the HT phase for ScV$_6$Sn$_6$.} Fermi surfaces including the hole pockets (a) and electron pockets(b). (c-d) The normalized imaginary (c) and real (d) parts of the electron susceptibility $\chi(q)$ on the $q_z=\frac{1}{3} c^*$. The green arrows denote $q_c=\frac{1}{3} a^*+\frac{1}{3} b^*+\frac{1}{3} c^*$, where $a^*, b^*, c^*$ are the reciprocal lattice vectors.
}\label{Fig:6}
\end{figure*}

To identify whether the FSN plays a role in the formation of CDW, we calculated the Fermi surfaces and the electron susceptibility of the HT phase. The hole pocket in  Fig.\ref{Fig:6}(a) consists of an open Fermi surface wrapping A and M inside and an ellipsoid surrounding K, while the electron pocket in Fig.\ref{Fig:6}(b) is a close one centering at M. The modulation wave vector of ScV$_6$Sn$_6$ is reported to be $(\frac{1}{3},\frac{1}{3},\frac{1}{3})$ \cite{Arachchige2022}. As the FSN can be reflected by the electron susceptibility $\chi{\boldsymbol(q)}$ \cite{Johannes2008}, we calculate the real part  $\chi_0^{\prime}{\boldsymbol(q)}$ defined as

\begin{equation}
\lim _{\omega \rightarrow 0} \chi_0^{\prime}(\boldsymbol{q})=\sum_{\boldsymbol{k}} \frac{f\left(\varepsilon_{\boldsymbol{k}}\right)-f\left(\varepsilon_{\boldsymbol{k}+q}\right)}{\varepsilon_{\boldsymbol{k}}-\varepsilon_{\boldsymbol{k}+\boldsymbol{q}}}
\end{equation}
and the imaginary part $\chi_0^{\prime\prime}{\boldsymbol(q)}$
\begin{equation} 
\lim _{\omega \rightarrow 0} \chi_0^{\prime \prime}(\boldsymbol{q}, \omega)=\sum_k \delta\left(\varepsilon_k-\varepsilon_F\right) \delta\left(\varepsilon_{k+q}-\varepsilon_F\right)
\end{equation}
on the $q_z=\frac{1}{3} c^*$ plane as shown in Fig.\ref{Fig:6}(c-d), which have been normalized for better illustration. It can be found that the maxima in Fig.\ref{Fig:6}(c-d) are both concentrated around $\Gamma$ and away from the wave vector indicated by the green arrow. Therefore, the CDW in ScV$_6$Sn$_6$ is not likely driven by the FSN.

It is inspiring to compare the optical spectroscopy results on ScV$_6$Sn$_6$ with those of the similar vanadium structural motif kagome metals AV$_3$Sb$_5$ (A=K,Rb,Cs). The opening of the CDW gap behavior is clearly observed in AV$_3$Sb$_5$ below T$_{cdw}$ in previous optical reflectivity studies\cite{Zhou2021,Uykur2021,Uykur2022,Wenzel2022a}. 
The CDW gap formation results in a transfer of spectral weight from low to higher energy regions, causing a reduction in the weight of the Drude component. The importance of saddle point nesting in driving the CDW instability in CsV$_3$Sb$_5$ has been proposed. 
Some other experiment techniques have also revealed first-order phase transition features, such as a sudden change in the ultrafast relaxation dynamics\cite{Wang2021} and the absence of a CDW amplitude mode\cite{Wang2021,Liu2022}.
Thus, both first and second order phase transition related behaviors have been observed in AV$_3$Sb$_5$. But ScV$_6$Sn$_6$ displays different behaviors, exhibiting only a sudden band change behavior that is closely related to the structural transition. The optical spectra overlap with almost imperceptible differences below T$_s$. Therefore, this behavior is attributed to a pure first-order structural transition similar to those observed in BaNi$_2$As$_2$\cite{PhysRevB.80.094506}, RuP\cite{Chen2015}, IrTe$_2$\cite{Fang2013} and TaTe$_2$\cite{Hu2022}. 

Optical study can provide critical information about the origin of the structural transition\cite{Huang2013,Huang2014,Wang2014}. Based on the above experiment and calculations results, it is almost certain that the conventional FSN is not the driving mechanism for the structural phase transition in ScV$_6$Sn$_6$. It has been calculated to have some exotic saddle points near the Fermi surface in ScV$_6$Sn$_6$. One possible hypothesis is that these saddle points close to the Fermi energy result in a sudden lattice instability and CDW emergence at low temperatures. However, more  work is necessary to fully understand the driving nature of the structural phase transition in ScV$_6$Sn$_6$ detailedly.

\section{SUMMARY}
In conclusion, our study of a single crystal of ScV$_6$Sn$_6$ has revealed that it undergoes a first-order phase transition and remains metallic. The optical spectra showed evidence of sudden changes in band structure, which was confirmed by our calculations. Despite some reduction of free carriers across the transition, the decrease in scattering rates improved the metallic properties of the LT phase. There was no gradual gap-opening  observed in spectra, indicating the phase transition is of the first-order type and is irrelevant to the conventional CDW instability. Comparing with the widely-studied motif AV$_3$Sb$_5$ (A= K,Rb,Cs) series, the nature of the phase transition seems to be different. Our findings on ScV$_6$Sn$_6$ are illuminating for further investigation into the origin of this intriguing phase transition in the kagome metal.

\begin{center}
\small{\textbf{ACKNOWLEDGMENTS}}
\end{center}

This work was supported by the National Natural Science Foundation of China (Grants No.11888101, No.11925408, No.11921004 and No.12188101), the National Key Research and Development Program of China (Grants No.2018YFA0305700, No.2022YFA1403800 and 2022YFA1403900), the Chinese Academy of Sciences (Grant No.XDB33000000) and the Informatization Plan of the Chinese Academy of Sciences (Grant No.CAS-WX2021SF-0102), and the Center for Materials Genome.

\bibliography{ScVSn}

\end{document}